# Scaling of Magnetic Domain Walls in Perpendicular Magnetic Anisotropy Systems

Guowen Gong[1,2], Changmin Xiong[3,4], Lijun Zhu[1,2]*

1. State Key Laboratory of Semiconductor Physics and Chip Technologies, Institute of Semiconductors, Chinese Academy of Sciences, Beijing 100083, China
2. College of Materials Science and Opto-Electronic Technology, University of Chinese Academy of Sciences, Beijing 100049, China.
3. Department of Physics, Beijing Normal University, Beijing 100875, China
4. Key Laboratory of Multiscale Spin Physics, Ministry of Education, Beijing 100875, China

*ljzhu@semi.ac.cn

**Abstract**: Magnetic domain walls play a critical role in the nanoscale evolution of magnetic devices. Despite the early efforts, a complete understanding of the micromagnetic evolution of the width (Δ) and the type of magnetic domain walls has still remained missing. Here, we report a combined analytical and micromagnetic simulation study and establish the scaling of the magnetic domains as a function of the exchange stiffness ($A$), uniaxial perpendicular magnetic anisotropy ($K_u$), saturation magnetization ($M_s$), and Dzyaloshinskii-Moriya interaction (DMI), and shape anisotropy of the magnetic device. We find that Δ of both Bloch and Néel walls scales excellently with the analytical prediction of $\Delta = C\sqrt{A/[K_u - \mu_0 M_s^2(N_z - N_x\cos^2\varphi)/2]}$ (where $C$ is a constant depending on the definition of Δ, $\varphi$ is the azimuth angle of the magnetic moment at the domain wall center, $N_x$ and $N_z$ are the longitudinal and perpendicular demagnetization factors). The DMI is found to have little influence on the domain wall width but strongly affect the type of the domain wall. The domain wall has a Bloch configuration at zero DMI and gradually transitions to Néel configuration upon increase of the DMI. The shape anisotropy of the magnetic domain wall also affects the domain wall width. These results have established a comprehensive, conclusive understanding of the magnetic domain walls within spintronics devices.

## I. Introduction

Manipulation of nanomagnet with perpendicular magnetic anisotropy (PMA) has enabled the development of a variety of high-performance spintronic devices, ranging from memories, computing, and sensors[1-8]. Microscopically, the functionality of these spintronic devices is achieved via the nanoscale evolution of magnetic domains.[9,10] Based on the spin torque-driven displacement of magnetic domain walls (DWs), racetrack memory[11], nanowire logic[12], and magnetic tunnel junctions[13-16] have been demonstrated utilizing ferromagnets and ferrimagnets. The type of magnetic DWs is critical for the effectiveness of the spin torque manipulation of such devices, with efficient manipulation expected for Néel-type DWs but not Bloch-type ones[17]. The width (Δ) of a magnetic DW typically affects the displacement velocity and Walker breakdown current density of the DWs in a linear manner in the flow region[18,19], and thus the operation speed of the device. Magnetic DWs can mediate chiral coupling of adjacent orthogonal magnetic domains[4] and switch traveling magnetic domains[12] and magnons[20,21].

Despite the early efforts[18,19,22-27] and the fundamental importance of magnetic DWs for spintronics, quantitative understanding of the evolution of the magnetic domain walls, including the type and the width, have remained largely missing. Experimental investigation of the magnetic domain walls of PMA devices has remained challenging. The domain walls of PMA devices are predicted to be typically of a few nanometers in width[18,26,27,28,29], which is below the resolutions of available experimental techniques, such as nitrogen-vacancy magnetometry[30,31], magneto-optical Kerr microscopy[4,32], Lorentz transmission electron microscopy[33,34], and magnetic force microscopy[35,36]. In addition, it remains quantitatively unclear as to how the DMI[37-39] drives the transition of the DW from the achiral Bloch type to the chiral Néel type and as to whether the DMI affects the domain wall width. In the literature reports, the demagnetization factor of the domain wall has also been either ignored[40] or approximated differently in the estimation of the influence on the DW width[18,19,26,41], leaving the role of the demagnetization factor an open question.

In this work, we perform a systematic study of the evolution of the type and width of the magnetic domain walls in magnetic films with perpendicular magnetic anisotropy as a function of the exchange stiffness ($A$), the uniaxial perpendicular magnetic anisotropy ($K_u$), the saturation magnetization ($M_s$), the DMI strength ($D$), and the demagnetization factor ($N_x$) by combining both the analytical solution and micromagnetic simulations.

## II. Analytical derivation of the domain wall width

To quantify the width of the DW, we first analytically derive the equation for an arbitrary DW by considering a continuous spatial rotation of the magnetic moments within the DW. Taking into account the exchange energy, uniaxial PMA, shape anisotropy, and DMI energy, the total energy density of the DW is given by

$$E = \int [A(\frac{\partial\theta}{\partial x})^2 + K_u\sin^2\theta - \frac{1}{2}\mu_0 M_s^2 \sin^2\theta (N_z - N_x\cos^2\varphi) - \pi D\cos\varphi/\Delta]dx, \quad (1)$$

under the continuous assumption.[42] As shown in Fig. 1(a), $x$ and $y$ are directions perpendicular and parallel to the domain wall, $z$ is the film normal direction, $\theta$ and $\varphi$ are the polar and azimuth angles of the magnetic moment, $N_x$ and $N_z$ are the demagnetization factors of the magnetic film in the $x$ and $z$ directions, respectively. Here, we have ignored the term of the transverse demagnetization factor, $-N_y\sin^2\varphi$, because the magnetic DWs considered here typically have much greater transverse dimension (~100 nm) than the DW width (~ a few nm)[18,19,26,41] and the layer thickness (~ 1 nm).

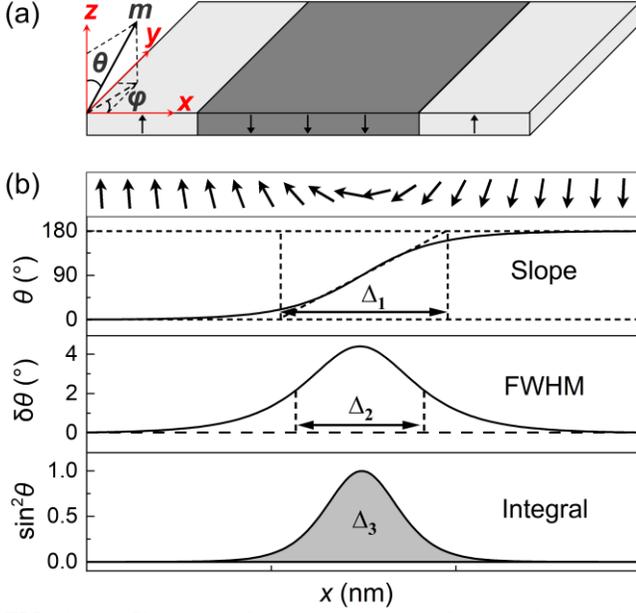

FIG. 1. (a) Sketch of the initial state of the simulation. (b) Distribution and the polar angle ($\theta$) of the magnetic moment within a stable domain wall state, the relative orientation of the two adjacent magnetic moments ($\delta\theta$), and $\sin^2\theta$ as a function of $x$ position (the data are from micromagnetic simulations with $A = 3\times10^{-11}$ J/m, $K_u = 5\times10^6$ J/m³, $M_s = 1\times10^6$ A/m, and $D = 1$ mJ/m²).

The variational method analysis with $\frac{\partial E}{\partial \varphi} = 0$ yields

$$\varphi = \begin{cases} \arccos\frac{\pi D}{2\mu_0 M_s^2 N_x \Delta} & \text{for } 0 \leq |D| \leq |D_c| \\ 0° & \text{for } |D| > |D_c|, \end{cases} \quad (2)$$

with the critical DMI constant of $D_c = \frac{2\mu_0 M_s^2 N_x \Delta}{\pi}$. The domain wall is expected to transition from Bloch-type ($\varphi = 90°$ at the center of the domain wall $x_0$) to Néel-type ($\varphi = 0°$) as $|D|$ increases from zero to $|D_c|$, and becomes energetically stable in the complete Néel-type configuration for $|D| > |D_c|$. According to the variational method and the energy minimization analysis of the energy density $E$ as a function of the position $x$, one gets the spatial distribution and the polar angle ($\theta$) of the magnetic moment within a stable domain wall (Fig. 1b), i.e.,

$$\theta(x) = 2\arctan(e^{\frac{x-x_0}{\sqrt{A/K_{\text{eff}}}}}), \quad (3)$$

$$\frac{\partial x}{\partial \theta} = \frac{1}{2}\sqrt{A/K_{\text{eff}}}\,[\sec^2(\theta/2)/\tan(\theta/2)], \quad (4)$$

$$K_{\text{eff}} = K_u - \mu_0 M_s^2 (N_z - N_x \cos^2\varphi)/2. \quad (5)$$

Below we consider three definitions of the DW width. In the simplest slope model,[43] $\Delta$ is defined as $\pi$ times the first position derivate of the moment orientation ($\frac{\partial x}{\partial \theta}|_{\theta=90°}$) at the center of the DW, i.e.,

$$\Delta_1 = \pi \frac{\partial x}{\partial \theta}|_{\theta=90°} = \pi\sqrt{A/K_{\text{eff}}}. \quad (6)$$

In the second model, we define the DW width as the "full width at half maximum (FWHM)" of the spatial distribution peak of the relative orientation of the adjacent magnetic moments ($\delta\theta$). $\delta\theta$ is zero at the center of each magnetic domain and gradually increases as the position $x$ goes towards the center of the magnetic DW in between the associated domains. The analytical result of the FWHM of the $\delta\theta$ peak is written as

$$\Delta_2 = 2\ln(2+\sqrt{3})\sqrt{A/K_{\text{eff}}} \approx 2.63\sqrt{A/K_{\text{eff}}}. \quad (7)$$

In the third integral model[44] we integrate the spatial distribution of all magnetic moments within the domain wall, i.e.,

$$\Delta_3 = \int \sin^2\theta\, dx = 2\sqrt{A/K_{\text{eff}}}. \quad (8)$$

As a result, the width of the DW can be written in the unified form of

$$\Delta = C\sqrt{A/[K_u - \mu_0 M_s^2(N_z - N_x\cos^2\varphi)/2]}, \quad (9)$$

with $C$ being $\pi$, 2.63, and 2 for the slope, FWHM, and integral models, respectively.

For a Bloch wall with $\varphi = 90°$, the demagnetization factor $N_x$ plays no role. For a Néel wall ($\varphi = 0°$) or a mixed wall configuration ($0° < \varphi < 90°$), $N_x$ can play a critical role. In the literature, $N_x$ has been assumed zero,[40] estimated as $t/(t+\Delta_0)$ following an ellipsoid approximation[18,19,26] and as $t\ln 2/\pi\Delta_0 - t^2/18\Delta_0^2$ by more explicitly incorporating nonuniform magnetization within the domain wall[41]. Here, $\Delta_0 \equiv \Delta/C$. In the fits of the micromagnetic simulation results of the DW width below, we employ $N_x = t\ln 2/\pi\Delta_0 - t^2/18\Delta_0^2$ for Eqs. (6)-(9) unless otherwise mentioned.

**III. Micromagnetic simulations**

To verify the accuracy of the analytical derivations of the width and the types of the domain wall, we perform systematic micromagnetic simulations using OOMMF on a 400 nm×200 nm×1 nm uniaxial PMA magnetic device with a very small unit cell size of 0.2 nm×0.2 nm×0.2 nm. During the simulations, the device is first set to the initial state with three magnetic domains separated by two perpendicular DWs along the $x$ direction (Fig. 1a) and then spontaneously relax to the minimum energy state.

We first discuss the scaling of the DW width with the exchange stiffness $A$. From the cross-sectional images of the simulated DW configurations in Figs. 2a,b, the DW is stabilized in a Bloch configuration with spins rotating in the $yz$ plane in the absence of the DMI ($D = 0$ mJ/m², Fig. 2a) but in a Néel configuration with spins rotating in $xz$ plane for a sufficiently strong DMI ($D = 1$ mJ/m², Fig. 2b). The domain wall region expands quickly as $A$ increases, regardless the Bloch or Néel configuration. More quantitatively, the micromagnetic simulation results of the DW width ($K_u = 5\times10^6$ J/m³ and $M_s = 1\times10^6$ A/m, Figs. 2c,d) always increase as $A$ increases, regardless the three definitions, the scaling of which are fit very well by Eq. (9). As shown in Figs. 2e,f, the relative errors, defined as the ratio of the difference between the analytical equations and the simulation results relative to the simulation results, are below 4% in the whole studied $A$ range, including the low-$A$ limit where the DW become atomically narrow.

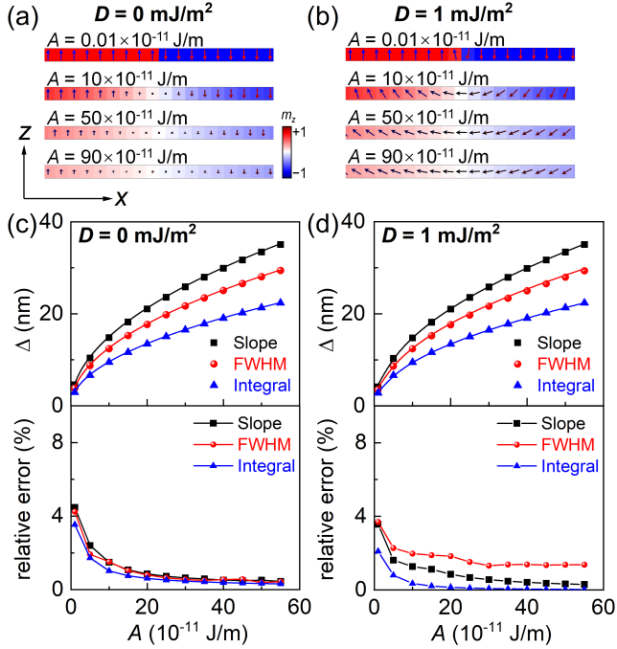

FIG. 2. Cross-sectional view of the magnetic domain wall with different $A$ for (a) Bloch wall ($D = 0$ mJ/m²) and (b) Néel wall ($D = 1$ mJ/m²). Dependences on $A$ of the DW width and the relative error for (c) Bloch wall ($D = 0$ mJ/m²) and (d) Néel wall ($D = 1$ mJ/m²) within the slope, FWHM, and integral models. The black, red, and blue lines plot the analytical results of the domain wall widths predicted by Eq. (6), Eq. (7), and Eq. (8), respectively. For data in (a)-(d), $K_u = 5 \times 10^6$ J/m³, $M_s = 1 \times 10^6$ A/m.

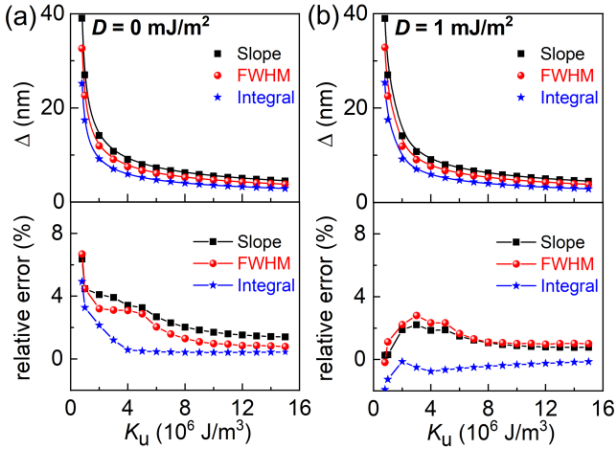

FIG. 3. Dependences on $K_u$ of the DW width and the relative error for (a) Bloch wall ($D = 0$ mJ/m²) and (b) Néel wall ($D = 1$ mJ/m²) within the slope, FWHM, and integral models ($A = 3 \times 10^{-11}$ J/m, $M_s = 1 \times 10^6$ A/m). The black, red, and blue lines plot the analytical results of the domain wall widths predicted by Eq. (6), Eq. (7), and Eq. (8), respectively.

In Figs. 3a,b we plot the simulated results of the DW width for both Bloch ($D = 0$ mJ/m²) and Néel walls (1 mJ/m²) as a function of $K_u$ for fixed $A = 3 \times 10^{-11}$ J/m and $M_s = 1 \times 10^6$ A/m. The DW width decreases rapidly as $K_u$ increases following the prediction of Eq. (9). We find that the relative error is very small (<4%) for the most cases except for the low-$K_u$ limit. In the low-$K_u$ limit, the device gradually losses PMA and transitions into wide in-plane domain walls due to the competing shape anisotropy of the thin film that favors in-plane anisotropy.

In Figs 4a,b, we plot the simulated results of the DW width for both Bloch ($D = 0$ mJ/m²) and Néel walls (1 mJ/m²) as a function of $M_s$ for fixed $A = 3 \times 10^{-11}$ J/m and $K_u = 5 \times 10^6$ J/m³. The scaling of $\Delta$ with $M_s$ can be fit by Eq. (9) in the entire realistic range (1.8×10⁶ A/m). The relative error is reasonably small although it increases slowly with $M_s$ because a very high $M_s$ would compete with PMA and cause distortion of 180° domain wall into in-plane.

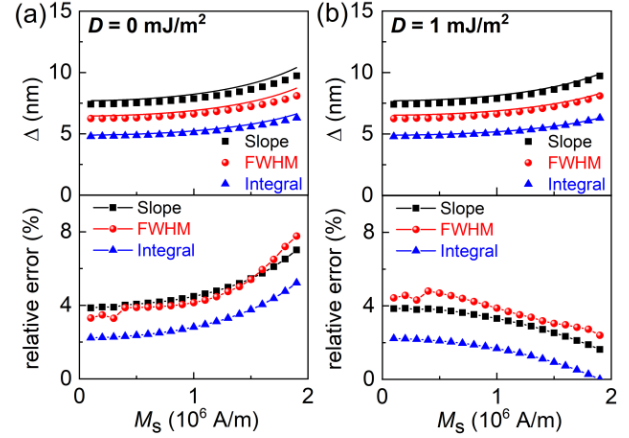

FIG. 4. Dependences on $M_s$ of the DW width and the relative error for (a) Bloch wall ($D = 0$ mJ/m²) and (b) Néel wall ($D = 1$ mJ/m²) within the slope, FWHM, and integral models ($A = 3 \times 10^{-11}$ J/m and $K_u = 5 \times 10^6$ J/m³). The black, red, and blue lines plot the analytical results of the domain wall widths predicted by Eq. (6), Eq. (7), and Eq. (8), respectively.

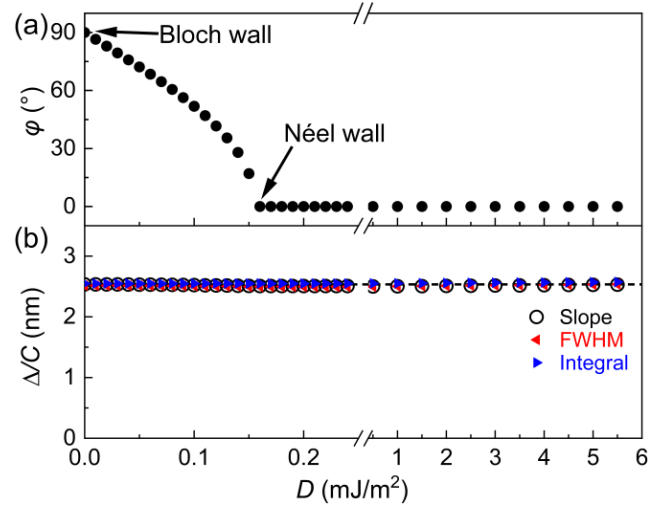

FIG. 5. (a) Azimuth angle $\varphi$ of the magnetic moment at the center of the DW plotted as a function of $D$. (b) Dependence on $D$ of the normalized DW widths $\Delta/C$ within the slope, FWHM, and integral models.

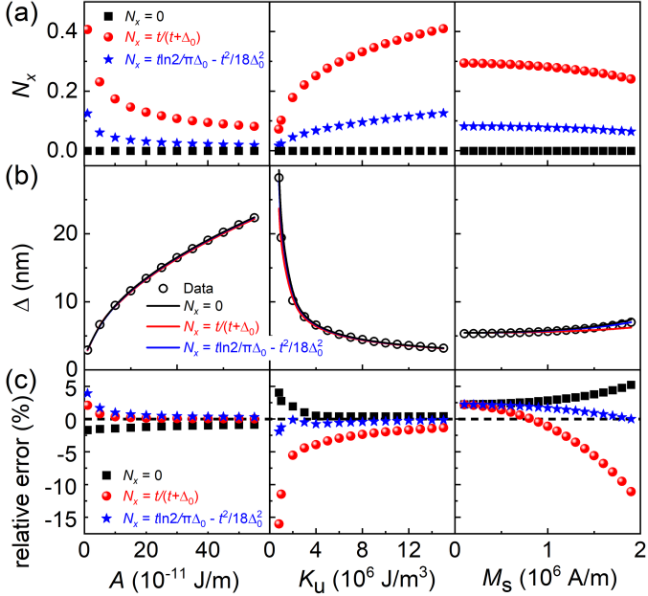

FIG. 6. (a) Estimated demagnetization factor of the domain wall ($N_x$) from $N_x = 0$, $t/(t+\Delta_0)$, and $t\ln2/\pi\Delta_0 - t^2/18\Delta_0^2$. (b) Micromagnetic simulations results (circle) and the analytical results of the domain wall width (the integral model) plotted as a function of $A$ ($D = 1$ mJ/m², $K_u = 5\times10^6$ J/m³, $M_s = 1\times10^6$ A/m), $K_u$ ($D = 1$ mJ/m², $A = 3\times10^{-11}$ J/m, $M_s = 1\times10^6$ A/m), and $M_s$ ($D = 1$ mJ/m², $K_u = 5\times10^6$ J/m³, $A = 3\times10^{-11}$ J/m). The black, red, and blue lines plot the analytical results predicted by Eq. (9) with the different $N_x$ values in (a), respectively. (c) Error of the analytical results relative to the micromagnetic simulation results plotted as a function of $A$, $K_u$ and $M_s$.

Below we discuss the effect of the DMI on the type and the width of the DWs. We find that the DMI strongly affects the type of the magnetic DW. In Fig. 5a, we plot the representative results of the dependence on $D$ of the azimuth angle $\varphi$ of the magnetic moment at the center of the domain wall of a device with $A = 3\times10^{-11}$ J/m, $K_u = 5\times10^6$ J/m³, and $M_s = 1\times10^6$ A/m. The domain wall has a Bloch configuration ($\varphi = 0°$) at $D = 0$, which is consistent with the cross-sectional image in Fig. 2a. As $D$ increases, the domain wall transitions to a mixed Bloch and Néel configuration ($0° < \varphi < 90°$) and finally into a pure Néel configuration ($\varphi = 90°$) when the DMI exceeds a critical value (0.16 m J/m² for the material parameters in Fig. 5a), which is consistent with the prediction of Eq. (2). Our observation of the Bloch-Néel transition is supported by a previous micromagnetic simulation[30]. In contrast, the simulated DW width exhibits little variation in the wide range of the DMI constant (Fig. 5b). This observation also agrees well with the DW width data in Figs. 2-5.

Finally, we test the influence of the demagnetization factor $N_x$ on the domain wall with of the Néel wall. In Fig. 6a, we plot the values of $N_x$ as a function of $A$, $K_u$ and $M_s$. $N_x$ calculated using $t\ln2/\pi\Delta_0 - t^2/18\Delta_0^2$ varies between 0 and 0.15, but is always several times smaller than those predicted by $t/(t+\Delta_0)$. In Figs. 6b, we plot the simulated data of the DW width (the integral model) and the fits of the data to Eq. (9) using three different sets of the $N_x$ values in Fig. 6a. From the relative errors plotted in Fig. 6c, one can see that, overall, Eq. (9) with $N_x = t\ln2/\pi\Delta_0 - t^2/18\Delta_0^2$ best fits the data as a function of $A$, $K_u$ and $M_s$ with relative errors of below 4%. Ignoring the in-plane demagnetizing factor (set $N_x = 0$) only cause a deviation of below 7.5%. In contrast, Eq. (9) with $N_x = t/(t+\Delta_0)$ significantly losses accuracy at small $K_u$ and high $M_s$ due to overestimation of the $N_x$ by the ellipsoid approximation. Therefore, Eq. (9) with $N_x = t\ln2/\pi\Delta_0 - t^2/18\Delta_0^2$ provides the most ideal description of the linewidth, compared to the widely employed $N_x = 0$ and $N_x = t/(t+\Delta_0)$.

## V. CONCLUSIONS

We have presented a systematic study of the magnetic domains of magnetic films as a function of the key relevant parameters, including the exchange stiffness, uniaxial perpendicular magnetic anisotropy, saturation magnetization, and the DMI, and the in-plane demagnetization factor. We derived that analytical relation of the domain wall width and type of both the Bloch and Néel walls and verified the accuracy using micromagnetic simulations. We find that the analytical expression of Eq. (9) with $N_x = t\ln2/\pi\Delta_0 - t^2/18\Delta_0^2$ provides an ideal prediction of the width of the domain walls. While the DMI has a very weak influence on the domain wall width but strongly affect the type of the domain wall. The domain wall has a Bloch configuration at zero DMI and gradually transitions to Néel configuration upon increase of the DMI. These results have established a comprehensive and conclusive understanding of the type and width of magnetic domain walls, which are critical information for the micromagnetic configuration, velocity, Walker breakdown current density, domain wall-mediated chiral coupling of orthogonal magnetic domains[4], domain wall-domain coupling[12], domain wall-magnon coupling[20,21] within racetrack memory, nanowire logic, and magnetic tunnel junctions, and other spintronics devices.

**ACKNOWLEDGMENTS**

This work was supported partly by the National Key Research and Development Program of China (2022YFA1204000), by the Beijing Natural Science Foundation (Z230006), and by the National Natural Science Foundation of China (12274405).